\documentclass[journal,onecolumn]{IEEEtran}
\IEEEoverridecommandlockouts
\pdfoutput=1
\usepackage{amsmath}
\usepackage{cases}
\usepackage{graphicx}
\usepackage{textcomp}
\usepackage{color}
\usepackage{amsmath,amssymb,amsfonts}
\usepackage{xcolor} 
\usepackage{soul} 

\soulregister\cite7 % 针对\cite命令
\soulregister\citep7 % 针对\citep命令
\soulregister\citet7 % 针对\citet命令
\soulregister\ref7 % 针对\ref命令
\soulregister\pageref7 % 针对\pageref命令

\usepackage[american]{babel}%遇单词换行的时候，确保单词的切割是按照音节来而不是随意切割。
\usepackage{microtype}%调整全篇文章（或局部）的字间距，字间距最大调整范围为±1em。可使得某段落不会出现单独一个单词占一行，或文章末尾单独一行文字占一页的不美观情况
\usepackage{graphicx}%插入图片
\usepackage[ruled]{algorithm2e}
   \usepackage{algorithmicx}
\usepackage{setspace}
\usepackage{cite}
\usepackage{tcolorbox}

\usepackage{mathptmx}%把数学字体也改成类似Times的字体。这个字体真的不需要再多说什么了，总之我觉得看久了眼睛会累，但是打印的效果非常稳妥。

\usepackage{mathbbol}%一些特殊的数学公式符号？

\usepackage{caption}
\usepackage{framed}

\usepackage{subfigure}

\usepackage{amsmath}%公式split环境

\usepackage[numbers]{natbib}%使参考文献不带中括号
\setcitestyle{open={},close={}}

\usepackage{stfloats}%双栏中插入单栏的长公式

\usepackage{color}%自定义颜色

\usepackage{amsthm}%证明proof环境

\def\BibTeX{{\rm B\kern-.05em{\sc i\kern-.025em b}\kern-.08em
    T\kern-.1667em\lower.7ex\hbox{E}\kern-.125emX}}
\begin{document}
\title{Robust Beamforming for Enhancing Security
in Multibeam Satellite Systems}
\author{Jian~Zhang,
	Min~Lin,
	Jian Ouyang,
	Wei-Ping~Zhu,
	and ~Tomaso de~Cola,
	\thanks{J. Zhang, M. Lin  and J. Ouyang are with the College of Telecommunications and Information Engineering,
		Nanjing University of Posts and Telecommunications,
		Nanjing, China. 
		
		W. -P. Zhu is with the Department of Electrical and Computer Engineering, Concordia University, Montreal, Canada, and also with the College of Telecommunications and Information Engineering,
		Nanjing University of Posts and Telecommunications,
		Nanjing, China.
		
		T. de Cola is with the Institute of Communications and Navigation, German Aerospace Center (DLR), 82234 Oberpfaffenhofen, Germany.
		
		Corresponding author: Min Lin (linmin@njupt.edu.cn).
		
		This work was supported in part by the Key International Cooperation Research Project under Grant 61720106003, in part by the National Natural Science Foundation of China under Grant 61801234, in part by the Shanghai Aerospace Science and Technology Innovation Foundation under Grant SAST2019-095, and in part by NUPTSF under Grant NY220111. }

}
\maketitle
\captionsetup{belowskip=-10pt}

\begin{abstract}
     This paper proposes a robust beamforming (BF) scheme to enhance physical layer security (PLS) of the downlink of a multibeam satellite system in the presence of either uncoordinated or coordinated eavesdroppers (Eves). Specifically, with knowing only the approximate locations of the Eves, we aim at maximizing the worst-case achievable secrecy rate (ASR) of the legitimate user (LU), subject to the constraints of per-antenna  transmit power and quality of service (QoS) requirement of the LU. Since the optimization problem is non-convex, we first adopt the discretization method to deal with the unknown regions of the Eves and then exploit the log-sum-exp function to approximate the objective function. Afterwards, a BF method joint alternating direction method of multipliers (ADMM) with Dinkelbach iteration is presented to solve this non-convex problem. Finally, simulation results verify that our robust BF algorithm can effectively improve the security of multibeam satellite systems.
\end{abstract}

\begin{IEEEkeywords}
Multibeam satellite communications, robust beamforming, physical layer security, Alternating Direction Method of Multipliers
\end{IEEEkeywords}

\section{Introduction}
\IEEEPARstart
{T}{hanks} to the ability of offering high date rate transmission and achieving spectrum reuse, multibeam technology is considered as an indispensable element in the future satellite communication (SatCom) systems. [\cite{references1}]-[\cite{references4}]. However, the inherent nature of broadcast and wide coverage makes SatCom vulnerable to wiretapping. Therefore, security is regarded as a key issue in SatCom that has been widely studied in the open literatures. Although secure communication is traditionally achieved through cryptographic encryption at the upper layer, this approach faces great challenges with the emergence of new technologies such as cloud computing and quantum computing.

In recent years, physical layer security (PLS) technology has become an active research topic in  wireless communications [\cite{references5}]-[\cite{references6}]. Meanwhile,  beamforming (BF) technology can enhance the received signal power at legitimate users (LU) and suppress the signal leaked to the unintended users, and therefore  is an effective method to improve the security performance of wireless systems. Nowadays, the BF-based PLS technology has been widely investigated in the terrestrial cellular networks [\cite{references7}]-[\cite{references8}] and SatCom networks [\cite{references3}],[\cite{references9}]-[\cite{references10}]. Specifically, the application of PLS in SatCom systems  was studied in [\cite{references9}], where a sub-optimal BF algorithm based on zero forcing  and an optimal BF method using semi-definite programming relaxation (SDR) were proposed, respectively. In [\cite{references10}], the authors proposed a BF algorithm to improve the security performance of SatCom through green source of terrestrial networks. All the previous works  [\cite{references9}]-[\cite{references10}] verified that the PLS technology based on BF can effectively improve the security performance of the SatCom systems. However, the perfect channel state information (CSI) of eavesdroppers (Eves) is assumed to be known at the transmitter. In actual, this assumption does not hold since Eves are usually not authorized users [\cite{references11}]. Assuming that only the imperfect CSI of the wiretap channels is available, the authors in [\cite{references3}] proposed a robust BF scheme to maximize the minimal achievable secrecy rate (ASR) of LU under the constraint of  total transmit power on the satellite. However, it should be mentioned that the work in [\cite{references3}] only focused on the scenario of a single Eve, which is only a special case. Although the PLS problem could be solved by converting it to the form of SDR or second-order cone program (SOCP), there is still an urgent demand for developing computationally more efficient algorithms. This observation motivates our work in this paper.

Unlike most of the existing works, we here consider a practical scenario where only the approximate locations of the Eves are available, and propose a robust BF scheme to enhance the security for multibeam satellite systems in the presence of Eves regardless of whether they are coordinated Eves (CE) or uncoordinated Eves (UE). We first formulate a constrained optimization problem to maximize the worst-case ASR of the LU, while satisfying the on-board per-antenna  transmit power constraint and the quality of service (QoS) requirement for LU. Since the objective function is non-convex, we adopt the discretization method to deal with the unknown regions of the Eves and then approximate it through the log-sum-exp function. Afterwards, an iteration algorithm joint Dinkelbach and Alternating Direction Method of Multipliers (ADMM)  is adopted for solving this non-convex optimization problem. Compared to the existing approaches, the proposed scheme is computationally more efficient as  each iteration step
comes up with a simple closed-form solution.

\emph{Notations}:
Bold uppercase and lowercase letters denote matrices and vectors, respectively. $(\cdot)^{H}$stands for the Hermitian transpose, ${C^{m \times n}}$ denotes the complex space of $m \times n$, $CN\left( {\mu ,{\sigma ^2}} \right)$ represents a complex Gaussian distribution with mean $\mu $ and covariance ${\sigma ^{\text{2}}}$ matrix, ${\mathbf{a}} \odot {\mathbf{b}}$ denotes Hadamard product of two vectors    ${\mathbf{a}}$ and ${\mathbf{b}}$, ${[x,0]^ + } = \max \{ x,0\} $, ${[{\mathbf{w}}]_n}$ represents the $n$-th component of ${\mathbf{w}}$, $\left|  \cdot  \right|$ and $\left\|  \cdot  \right\|$  denote absolute value and Euclidean norm of a vector.

\vspace{-0.2cm}
\section{System Model}

In this paper, we consider the PLS for the downlink transmission of a multibeam satellite system, where a multibeam geostationary orbit (GEO) satellite sends private signal to the LU, while $K$ Eves within the satellite coverage attempt to overhear the private signal. Here, the GEO satellite configure $N$ antennas to generate $M$ beams. Besides, we suppose that only the imperfect CSI of each Eves is available\footnote{In secure communications, the Eve is usually passive and keeps silent, and thus only imperfect CSI of Eve can be obtained. In satellite communications, the transmitter can estimate the channel from the approximate location of the Eve which can be obtained by satellite GPS [\cite{references11}].}, which is a more practical assumption than that has been taken in the related works  [\cite{references9}]-[\cite{references10}], where perfect CSI is exploited to carry out BF design for PLS.

Suppose that the multibeam satellite transmits signal $x\left( t \right)$ with normalized power to the LU after performing  BF with weight vector ${\mathbf{w}} \in {C^{N \times 1}}$. The received signals at the LU and $k$-th Eve are, respectively, given by
\begin{equation}\label{eq1}
 {{\bf{y}}_s} = {\bf{h}}_s^H{\bf{w}}x\left( t \right) + {n_s}\left( t \right),
\end{equation}
\begin{equation}\label{eq2}
 {{\bf{y}}_k} = {\bf{h}}_k^H{\bf{w}}x\left( t \right) + {n_k}(t),k = 1...K,
\end{equation}
where ${n_s}\left( t \right)$  and ${n_k}\left( t \right)$  represent additive Gaussian white noises (AWGN) with zero means and the variances $\sigma _i^2 = \kappa {B_i}{T_i}$, $i \in \left\{ {s,k} \right\}$ at LU and $k$-th Eve respectively, with $\kappa  = 1.38 \times {10^{ - 23}}J/K$  denoting Boltzmann constant, ${T_i}$  the noise temperature and  ${B_i}$ the noise bandwidth. Further, $\left\{ {{{\mathbf{h}}_s},{{\mathbf{h}}_k}} \right\} \in {\mathbb{C}^{N \times 1}}$ are the channel vector between satellite and the LU, and that between satellite and $k$-th Eve, respectively. The channel vector of the satellite downlink can be written as [\cite{references10}]
\begin{equation}\label{eq3}
  {{\bf{h}}_m} = \sqrt {{G_r}}  \odot {\bf{r}}_m^{ - \frac{1}{2}} \odot {\bf{b}}_m^{\frac{1}{2}} \odot {{\bf{\hat h}}_m},
\end{equation}
where  ${{\mathbf{r}}_m} \!\!=\!\! {\!\left[\! {{r_{m1}},{r_{m2}}, \!\ldots\! ,{r_{mN}}} \!\right]\!^T}$ represents a rain attenuation fading vector, and whose entries expressed in dB $r_{mn}^{{\text{dB}}} \!=\! 20{\log _{10}}\left( {{r_{mn}}} \right)$   follow a lognormal random distribution, with  $\mu $ and $\sigma _r^{}$  being the lognormal location and the scale parameter, respectively. Besides ${{\mathbf{b}}_m}\! =\! {\left[ {{b_{m1}},{b_{m2}}, \!\ldots\! ,{b_{mN}}} \right]^T}$  is the  $N \times 1$ beam gain vector, whose component ${b_{mn}}$  is expressed as 
\begin{equation}\label{eq4}
 {b_{mn}} = {b_{\max }}{\left( {\frac{{{J_1}\left( {{u_{mn}}} \right)}}{{2{u_{mn}}}} + 36\frac{{{J_3}\left( {{u_{mn}}} \right)}}{{u_{mn}^3}}} \right)^2},
\end{equation}
where ${b_{\max }}$  represents the maximal beam gain,  ${J_1}\left( {\cdot} \right)$ and ${J_3}\left( {\cdot} \right)$  are the first-kind Bessel function of order 1 and 3, respectively, and ${u_{mn}} = 2.07123\sin {\phi _{mn}}/\sin {\phi _{3{\rm{dB}}}}$  with ${\phi _{mn}}$  being the angle between the $m$-th user’s position and the $n$-th beam center with respect to the satellite, and ${\phi _{3{\rm{dB}}}}$  the half power beamwidth. In (3), ${{\bf{\hat h}}_m} = {\left[ {{{\hat h}_{m1}},{{\hat h}_{m2}},...,{{\hat h}_{mN}}} \right]^T}$  is the channel response vector whose elements can be written as
\begin{equation}\label{eq5}
{\hat h_{mn}} = \frac{c}{{4\pi {f_c}{d_{mn}}}}{e^{ - j\frac{{2\pi {f_c}}}{c}{d_{mn}}}},
\end{equation}
where ${\!f_c\!}$  denotes the carrier frequency and $\!c\!$  the light speed. The variable  ${d_{mn}}$ is the distance between the $m$-th user and the $n$-th satellite antenna. In addition, let $\!{G_r}\!$  be the antenna gain of the user, as defined by [\cite{references12}]
\begin{equation}\label{eq6}
{G_r}\left[ {{\text{dB}}} \right] = \left\{ {\begin{array}{*{20}{c}}
	{{G_{\max }}}&,&{{0^ \circ } < {\theta _m} < {1^ \circ }} \\ 
	{32 - 25\log {\theta _m}}&,&{{1^ \circ } < {\theta _m} < {{48}^ \circ }} \\ 
	{ - 10}&,&{{{48}^ \circ } < {\theta _m} < {{180}^ \circ }} 
	\end{array}} \right.,
\end{equation}
where ${G_{\max }}$  denotes the maximum antenna gain of the user, and ${\theta _m}$  is off-boresight angle. According to (\ref{eq1}) and (\ref{eq2}), the signal-to-noise ratio (SNR) of the LU and that of the $k$-th Eve are ${\gamma _s} = {{{{\left| {{\mathbf{h}}_s^H{\mathbf{w}}} \right|}^2}} \mathord{\left/
		{\vphantom {{{{\left| {{\mathbf{h}}_s^H{\mathbf{w}}} \right|}^2}} {\sigma _s^2}}} \right.
		\kern-\nulldelimiterspace} {\sigma _s^2}}$  and ${\gamma _k} = {{{{\left| {{\mathbf{h}}_k^H{\mathbf{w}}} \right|}^2}} \mathord{\left/
		{\vphantom {{{{\left| {{\mathbf{h}}_s^H{\mathbf{w}}} \right|}^2}} {\sigma _k^2}}} \right.
		\kern-\nulldelimiterspace} {\sigma _k^2}}$, respectively. Similar to [\cite{references11}], we consider two cases where $K$ Eves overhear the private signal uncooperatively and cooperatively, respectively. In the case of UE, the ASR of LU can be written as
\begin{equation}\label{eq7}
 {R_s} = \mathop {\min }\limits_{k \in \left\{ {\left. {1,...,K} \right\}} \right.} {\left[ {\left. {\left( {\left. {{{\log }_{\rm{2}}}\left( {\left. {1 + {\gamma _s}} \right)} \right. - {{\log }_{\rm{2}}}\left( {\left. {1 + {\gamma _k}} \right)} \right.} \right)} \right.,0} \right]} \right.^{\rm{ + }}}.
\end{equation}
On the other hand, the ASR of LU in the case of CE can be expressed as
\begin{equation}\label{eq8}
 {R_s} = {\left[ {\left. {\left( {\left. {{{\log }_{\rm{2}}}\left( {\left. {1 + {\gamma _s}} \right)} \right. - {{\log }_{\rm{2}}}\left( {\left. {1 + \sum\limits_{k = 1}^K {{\gamma _k}} } \right)} \right.} \right)} \right.,0} \right]} \right.^{\rm{ + }}}.
\end{equation}
To guarantee the PLS in SatCom systems, we aim at designing the BF weight vector to maximize the ASR of the LU, which makes
the ASR positive, and therefore the superscript "+" can be removed [\cite{references9}].
\vspace{-0.2cm}
\section{Proposed Beamformig Scheme}
In this paper, we take the CSI uncertainty of Eve into account, and propose a worst-case robust BF scheme to improve the security performance of the system, by considering the scenarios of UE and CE, respectively. We here consider that the satellite only knows the approximate locations of the Eves. In other words, each Eve is located in an approximate bounded by a rectangle and thus the   CSI of the Eve can be expressed as
\begin{equation}\label{eq9}
{\Psi _k} \!\!=\!\! \!\left\{\! {{{\bf{h}}_k}\left| {{x_k} \!\in\! \left[ {{x_{k,L}},{x_{k,U}}} \right],{y_k} \in \left[ {{y_{k,L}},{y_{k,U}}} \right]} \right.} \!\!\right\}\!\!,\left\{ {k \in 1, \!\cdot  \cdot  \cdot\! ,K} \right\},
\end{equation}
where ${x_{k,L}}$ and ${x_{k,U}}$ denote the lower and upper bounds of the x-coordinate of the $k$-th Eve, and ${y_{k,L}}$ and ${y_{k,U}}$ the lower and upper bounds of the y-coordinate.
\vspace{-1em}
\subsection{Uncooperative Eves}
In this subsection, we focus on the case of UE, according (\ref{eq7}) and (\ref{eq9}), the worst-case problem can be written as
\begin{equation}\label{eq10}
\begin{gathered}
 \mathop {\max }\limits_{\mathbf{w}} {\text{ }}\mathop {\min }\limits_k {\text{ }}\mathop {\min }\limits_{{{\mathbf{h}}_k} \!\in\! {\Psi _k}} {\text{ }}{\log _2}\!\!\left(\!\! {\left. {1 \!\!+\!\! \frac{{{{\left| {{\mathbf{h}}_s^H{\mathbf{w}}} \right|}^2}}}{{\sigma _s^2}}} \right)} \right. \!\!-\!\! {\log _2}\!\!\left(\!\! {1 \!+\! \frac{{{{\left| {{\mathbf{h}}_k^H{\mathbf{w}}} \right|}^2}}}{{\sigma _k^2}}} \!\!\right)\!\! \hfill \\
\quad\quad\quad\quad\quad {\text{                   s}}{\text{.t}}{\text{.  }}\frac{{{{\left| {{\mathbf{h}}_s^H{\mathbf{w}}} \right|}^2}}}{{\sigma _s^2}} \geqslant {\gamma _{th}} \hfill \\
 \quad\quad\quad\quad\quad\quad\quad{\text{                          }}\left| {{{[{\mathbf{w}}]}_n}} \right| = \sqrt p ,n = 1,...,N \hfill \\ 
 \end{gathered},
\end{equation}
where $p$  represents the transmit power of each antenna, and  ${\gamma _{th}}$ the QoS requirement for LU. Here, we assume that the transmit power of per-antenna is constant, which can be easily implemented with low-cost phase shifters [\cite{references13}].

By denoting ${{\mathbf{\tilde h}}_s} = {{{{\mathbf{h}}_s}} \mathord{\left/
		{\vphantom {{{{\mathbf{h}}_s}} {{\sigma _s}}}} \right.
		\kern-\nulldelimiterspace} {{\sigma _s}}}$, ${{\mathbf{\tilde h}}_k} = {{{{\mathbf{h}}_k}} \mathord{\left/
		{\vphantom {{{{\mathbf{h}}_k}} {{\sigma _k}}}} \right.
		\kern-\nulldelimiterspace} {{\sigma _k}}}$ and considering that  ${\log _2}(\cdot)$ is a monotonically increasing function, we can obtain an equivalent form of (\ref{eq10}) as given below 
\begin{subequations}\label{eq11}
\begin{align}
\mathop {\max }\limits_{\mathbf{w}} {\text{ }}\mathop {\min }\limits_k {\text{ }}\mathop {\min }\limits_{{{\mathbf{h}}_k} \in {\Psi _k}}& {\text{ }}\frac{{1 + {{\mathbf{w}}^H}{{{\mathbf{\tilde h}}}_s}{\mathbf{\tilde h}}_s^H{\mathbf{w}}}}{{1 + {{\mathbf{w}}^H}{{{\mathbf{\tilde h}}}_k}{\mathbf{\tilde h}}_k^H{\mathbf{w}}}}    \\
{\text{s}}{\text{.t}}{\text{.   }}&\;\;{{\mathbf{w}}^H}{{\mathbf{\tilde h}}_s}{\mathbf{\tilde h}}_s^H{\mathbf{w}} \geqslant {\gamma _{th}} \\
&\;\;\left| {{{[{\mathbf{w}}]}_n}} \right| = \sqrt p ,n = 1,...,N.
\end{align}
\end{subequations}
It can be observed that the optimization problem in (\ref{eq11}) is equivalent to the following alternative form
\begin{equation}\label{eq12}
\begin{gathered}
\mathop {\min }\limits_{\mathbf{w}} {\text{ }}\frac{{\mathop {\max }\limits_k {\text{ }}\mathop {\max }\limits_{{{\mathbf{h}}_k} \in {\Psi _k}} {\text{ }}\left( {1 + {{\mathbf{w}}^H}{{{\mathbf{\tilde h}}}_k}{\mathbf{\tilde h}}_k^H{\mathbf{w}}} \right)}}{{1 + {{\mathbf{w}}^H}{{{\mathbf{\tilde h}}}_s}{\mathbf{\tilde h}}_s^H{\mathbf{w}}}} \hfill \\
{\text{ s}}{\text{.t}}{\text{.   (11b)}},{\text{(11c)  }} \hfill \\ 
\end{gathered}.
\end{equation}
It is clear from (\ref{eq9}) that the unknown area of $k$-th Eve ${\Psi _k}$  is a continuous rectangular area which makes the original problem in (\ref{eq12}) nonconvex and mathematically intractable. To overcome this difficulty, we discretize the continuous region as
\begin{equation}\label{eq13}
x_k^{(i)} = {x_{k,L}} + i\Delta x,i = 0,...,{M_1} - 1,
\end{equation}
\begin{equation}\label{eq14}
y_k^{(i)} = {y_{k,L}} + j\Delta y,j = 0,...,{M_2} - 1,
\end{equation}
where $\Delta x = \left( {{x_{k,U}} - {x_{k,L}}} \right)/{M_1}$,  $\Delta y = \left( {{y_{k,U}} - {y_{k,L}}} \right)/{M_2}$. Suppose that the wiretap channel of the $k$-th Eve ${{\bf{h}}_k}$  belongs to a given discrete region as described below
\begin{equation}\label{eq15}
\begin{array}{l}
\begin{aligned}
{\Lambda _k} =& \left\{ {{\bf{h}}_k^{_{\left( {i,j} \right)}}\left| {x_k^{(i)} \!=\! {x_{k,L}} + i\Delta x,i \!=\! 0,...,{M_1} - 1,} \right.} \right.\\
&{\rm{                                  }}\left. {{\rm{ }}y_k^{(j)} \!=\!{y_{k,L}} \!+\! j\Delta y,j = 0,\!...\!,{M_2} \!-\! 1,k \in \left\{ {1,\!...\!,K} \right\}} \right\}
\end{aligned}
\end{array}.
\end{equation}
\vspace{-0.5em}
At this point, we can obtain
\begin{equation}\label{eq16}
\mathop {\max }\limits_k {\rm{ }}\mathop {\max }\limits_{{{\bf{h}}_k} \in {\Psi _k}} {\rm{ }}\!\left(\! {1 \!+\! {{\bf{w}}^H}{{{\bf{\tilde h}}}_k}{\bf{\tilde h}}_k^H{\bf{w}}} \right) \!=\! \mathop {\max }\limits_k {\rm{ }}\mathop {\max }\limits_{{{\bf{h}}_k} \in {\Lambda _k}} {\rm{ }}\left( {1 \!+\! {{\bf{w}}^H}{{{\bf{\tilde h}}}_k}{\bf{\tilde h}}_k^H{\bf{w}}} \right).
\end{equation}
Since the objective function of the optimization problem in (\ref{eq12}) is a form of min-max-max, to tackle this difficulty, we here exploit the log-sum-exp function to approximate the objective function. Hence, equation (\ref{eq16}) can be approximated as [\cite{references14}]
\begin{equation}\label{eq17}
\begin{aligned}
\mathop {\max }\limits_k {\text{ }}\mathop {\max }\limits_{{{\mathbf{h}}_k} \in {\Lambda _k}}& {\text{ }}\left( {1 + {{\mathbf{w}}^H}{{{\mathbf{\tilde h}}}_k}{\mathbf{\tilde h}}_k^H{\mathbf{w}}} \right) \\
&\!\approx\! {\beta ^{ - 1}}\ln \left( {\sum\limits_{k = 1}^K {\sum\limits_{i = 0}^{{M_1} \!-\! 1} {\sum\limits_{j = 0}^{{M_2} \!-\! 1} {{e^{\beta \!\left(\! {1 \!+\! {{\mathbf{w}}^H}{\mathbf{\tilde h}}_k^{\left( {i,j} \right)}{\mathbf{\tilde h}}{{_k^{\left( {i,j} \right)}}^H}{\mathbf{w}}} \!\right)\!}}} } } } \!\right)\!
\end{aligned},
\end{equation}
where $\beta $ is a parameter that regulates the degree of approximation, with the increase of $\beta $, the approximation error will gradually decrease. Substituting (\ref{eq17}) into the optimization problem (12) and it can be reformulated as
\begin{equation}\label{eq18}
\begin{gathered}
  \mathop {\min }\limits_{\mathbf{w}} {\text{ }}\frac{{{\beta ^{ - 1}}\ln \left( {\sum\limits_{k = 1}^K {\sum\limits_{i = 0}^{{M_1} - 1} {\sum\limits_{j = 0}^{{M_2} - 1} {{e^{\beta \left( {1 + {{\mathbf{w}}^H}{\mathbf{\tilde h}}_k^{\left( {i,j} \right)}{\mathbf{\tilde h}}{{_k^{\left( {i,j} \right)}}^H}{\mathbf{w}}} \right)}}} } } } \right)}}{{1 + {{\mathbf{w}}^H}{{{\mathbf{\tilde h}}}_s}{\mathbf{\tilde h}}_s^H{\mathbf{w}}}} \hfill \\
  {\text{   s}}{\text{.t}}{\text{.  (11b)}},{\text{(11c)  }} \hfill \\ 
\end{gathered}.
\end{equation}
Obviously, the objective function in (\ref{eq18}) is non-convex and in the form of fraction which can be efficiently solved by means of the  Dinkelbach method  [\cite{references15}]. The corresponding Dinkelbach sub-problem can then be  written as
\begin{equation}\label{eq19}
\begin{gathered}
  \mathop {\min }\limits_{\mathbf{w}} {\rm{ }}\Gamma ({\bf{w}}) \!\!=\!\! {\text{  }}{\!\beta\! ^{ \!\!- \!\!1}}\!\!\ln \sum\limits_{k \!=\! 1}^K {\sum\limits_{i \!=\! 0}^{{M_1} \!\!-\!\! 1} {\sum\limits_{j \!=\! 0}^{{M_2} \!\!-\!\! 1} {{e^{\beta \!\left(\! {1 \!+\! {{\mathbf{w}}^H}{\mathbf{\tilde h}}_k^{\!\left(\! {i,j} \!\right)\!}{\mathbf{\tilde h}}{{_k^{\!\left(\!\! {i,j} \!\!\right)\!}}^H}{\mathbf{w}}} \!\right)\!}}} } } \!\!\! - \!\!\!\eta \!\!\left(\!\! {1 \!\!+\!\! {{\mathbf{w}}^H}{{{\mathbf{\tilde h}}}_s}{\mathbf{\tilde h}}_s^H{\mathbf{w}}} \!\!\right)\!\! \hfill \\
  {\text{  s}}{\text{.t}}{\text{. (11b)}},{\text{(11c)  }} \hfill \\ 
\end{gathered}. 
\end{equation}
Since the problem in (\ref{eq19}) is still non-convex and difficult to solve, we here exploit a non-convex ADMM algorithm to solve it. First, variables ${\bf{\tilde w}}$  and  ${\bf{x}}$ are introduced to transform the above optimization problem to a suitable form for the ADMM solution framework, the problem in (\ref{eq19}) is equivalent to the following alternative form 
\begin{equation}\label{eq20}
\begin{gathered}
\mathop {\min }\limits_{{\mathbf{\tilde w}},{\mathbf{x}}} {\text{ }}\Gamma ({\mathbf{\tilde w}}) \hfill \\
\;\;{\text{s}}{\text{.t}}{\text{.    }}{{{\mathbf{\tilde w}}}^H}{{{\mathbf{\tilde h}}}_s}{\mathbf{\tilde h}}_s^H{\mathbf{\tilde w}} \geqslant {\gamma _{th}}, \hfill \\
\;\;\;\;\;\;\;{\text{        }}\left| {{{[{\mathbf{x}}]}_n}} \right| = \sqrt p ,n = 1,...,N, \hfill \\
\;\;\;\;\;\;\;\;{\text{         }}{\mathbf{x}} = {\mathbf{\tilde w}}. \hfill \\ 
\end{gathered} 
\end{equation}
Then, the augmented Lagrangian of (\ref{eq20}) can be written as
\begin{equation}\label{eq21}
L\left( {{\mathbf{\tilde w}},{\mathbf{x}},{\mathbf{v}}} \right) = \Gamma ({\mathbf{\tilde w}}) + \operatorname{Re} \left\{ {{{\mathbf{v}}^H}\left( {{\mathbf{\tilde w}} - {\mathbf{x}}} \right)} \right\} + \frac{\rho }{2}{\left\| {{\mathbf{\tilde w}} - {\mathbf{x}}} \right\|^2}
\end{equation}
Here,  ${\bf{v}}$ is a vector of Langrangian multiplier and $\rho $  the penalty factor. Finally, the variables  $\left\{ {\left. {{\bf{\tilde w}},{\bf{x}},{\bf{v}}} \right\}} \right.$ are cyclically updated
according to the non-convex ADMM algorithm framework.
Specifically, the following three steps are cycled through the
solution process
\begin{numcases}{}
{{\mathbf{x}}^{\left( {l \!\!+\!\! 1} \right)}} \!\!\!=\!\!\! {\text{ }}\mathop {\arg \min }\limits_{\left| {{{[{\mathbf{x}}]}_n}} \right| \!\!=\!\! \sqrt p } \operatorname{Re} \left\{ {{{\mathbf{v}}^{(l)}}^H\left( {{{{\mathbf{\tilde w}}}^{(l)}} \!\!- {\mathbf{x}}} \right)} \right\} \!\!+\!\! \frac{\rho }{2}{\left\| {{{{\mathbf{\tilde w}}}^{(l)}} -\!\! {\mathbf{x}}} \right\|^2}, \label{eq22}\\
\begin{aligned}\label{eq23}
{{\bf{\tilde w}}^{\left( {l + 1} \right)}} \!\!=\!\!& \mathop {\arg \min }\limits_{{{{\bf{\tilde w}}}^H}{{{\bf{\tilde h}}}_s}{\bf{\tilde h}}_s^H{\bf{\tilde w}} \ge {\gamma _{th}}} {\rm{ }}{\mathop{\rm Re}\nolimits} \left\{ {\nabla \Gamma {{\left( {{{\bf{x}}^{\left( {l + 1} \right)}}} \right)}^H}\left( {{\bf{\tilde w}} \!\!-\!\! {{\bf{x}}^{\left( {l + 1} \right)}}} \right)} \right\}\\
& \!\!+\!\! {\mathop{\rm Re}\nolimits} \left\{ {{{\bf{v}}^{(l)}}^H\left( {{\bf{\tilde w}} \!\!-\!\! {{\bf{x}}^{(l + 1)}}} \right)} \right\} \!\!+\!\! \frac{{\rho  + L}}{2}{\left\| {{\bf{\tilde w}} - {{\bf{x}}^{(l + 1)}}} \right\|^2},
\end{aligned}\\
{{\mathbf{v}}^{\left( {l + 1} \right)}} = {{\mathbf{v}}^{\left( l \right)}} + \rho \left( {{{\mathbf{\tilde w}}^{\left( {l + 1} \right)}} - {{{\mathbf{x}}}^{\left( {l + 1} \right)}}} \right).\label{eq24}
\end{numcases}
In (\ref{eq23}), $L$ is a constant greater than zero and satisfies $\left\| {\nabla \Gamma \left( {{{\bf{x}}_1}} \right) - \nabla \Gamma \left( {{{\bf{x}}_2}} \right)} \right\| \le L\left\| {{{\bf{x}}_1} - {{\bf{x}}_2}} \right\|,\forall {{\bf{x}}_1},{{\bf{x}}_{\rm{2}}}$. 

Next, let us elaborate on the solutions of ADMM sub-problems (\ref{eq22}) and (\ref{eq23}).

1)	Update the variable  ${\bf{x}}$

After some trivial computations, problem (\ref{eq22}) can be rewritten as
\begin{equation}\label{eq25}
\begin{gathered}
  \mathop {\min }\limits_{\mathbf{x}} {\text{ }}{\left\| {{\mathbf{x}} - \left( {{{{\mathbf{\tilde w}}}^{\left( l \right)}} + {\rho ^{ - 1}}{{\mathbf{v}}^{\left( l \right)}}} \right)} \right\|^2} \hfill \\
  \;{\text{s}}{\text{.t}}{\text{.    }}\;\left| {{{[{\mathbf{x}}]}_n}} \right| = \sqrt p ,n = 1,...,N \hfill \\ 
\end{gathered}. 
\end{equation}
It is clear that problem (\ref{eq25}) can be viewed as the Euclidean projection onto an Euclidean ball. The optimal solution is thus given by
\begin{equation}\label{eq26}
{[{{\mathbf{x}}^{(l{\text{ \!+\! 1}})}}]_n}{\text{ \!\!=\!\! }}\left\{ \begin{gathered}
\frac{{\sqrt p {{\left[ {{{{\mathbf{\tilde w}}}^{\left( l \right)}} \!+\! {\rho ^{ - 1}}{{\mathbf{v}}^{\left( l \right)}}} \right]}_n}}}{{\left| {{{\left[ {{{{\mathbf{\tilde w}}}^{\left( l \right)}} \!+\! {\rho ^{ \!-\! 1}}{{\mathbf{v}}^{\left( l \right)}}} \right]}_n}} \right|}},{\left[ {{{\mathbf{w}}^{\left( l \right)}} \!\!+\!\! {\rho ^{ \!\!-\!\! 1}}{{\mathbf{v}}^{\!\left(\! l \!\right)\!}}} \right]_n} \ne 0, \hfill \\
{[{{\mathbf{x}}^{(l)}}]_n},{\text{      }}else \hfill \\ 
\end{gathered}  \right.
\end{equation}

2)	Update the variable ${\bf{\tilde w}}$

Problem (\ref{eq23}) can be rewritten as
\begin{equation}\label{eq27}
\begin{gathered}
  \mathop {\min }\limits_{{\mathbf{\tilde w}}} {\text{ }}{\left\| {{\mathbf{\tilde w}} - \left( {{{\mathbf{x}}^{\left( {l + 1} \right)}} - \frac{{\nabla \Gamma \left( {{{\mathbf{x}}^{\left( {l + 1} \right)}}} \right) + {{\mathbf{v}}^{\left( l \right)}}}}{{\rho  + L}}} \right)} \right\|^2} \hfill \\
  \;\;{\text{s}}{\text{.t}}{\text{.   }}\;{{{\mathbf{\tilde w}}}^H}{{{\mathbf{\tilde h}}}_s}{\mathbf{\tilde h}}_s^H{\mathbf{\tilde w}} \geqslant {\gamma _{th}} \hfill \\ 
\end{gathered}. 
\end{equation}
It can be observed that (\ref{eq27}) defines a Quadratical Constraint Quadratic Programming problem with only one constraint, namely QCQP-1, which has a closed-form solution as discussed in [\cite{references16}].

Finally, the proposed algorithm is summarized as algorithm 1. The solution obtained by this algorithm is suboptimal, but the optimal solution can be obtained by arbitrarily reducing the iteration parameter $\varepsilon$.  The major complexity comes from ADMM iterations. And the total computational complexity of  each ADMM iteration is $O\left( {K{M_1}{M_2}\left( {8{N^2} + 11N + 5} \right) + 16{N^2} + 27N} \right)$.
\vspace{-0.2cm}
\subsection{Cooperative Eves}
\vspace{-0.1em}
In this subsection, we extend the proposed algorithm to the case of CE. In this regard, the constrained optimization problem can be formulated as the following worst-case problem
\begin{equation}\label{eq28}
\begin{gathered}
  \mathop {\!\max\! }\limits_{\mathbf{w}} {\text{ }}\mathop {\!\min\! }\limits_{{{\mathbf{h}}_k} \!\in\! {\Psi _k}} {\text{  }}{\!\log _2\!}\!\!\left(\!\! {\left. {1 \!\!+\!\! \frac{{{{\left| {{\mathbf{h}}_s^H{\mathbf{w}}} \right|}^2}}}{{\sigma _s^2}}} \!\!\right)\!\!} \right. \!-\! {\!\log _2\!}\!\left(\! {1 \!\!+\!\! \sum\limits_{k \!=\! 1}^K {\left( {\frac{{{{\left| {{\mathbf{h}}_k^H{\mathbf{w}}} \right|}^2}}}{{\sigma _k^2}}} \right)} } \!\right)\! \hfill \\
  \quad\quad\quad{\text{           s}}{\text{.t}}{\text{.   }}\frac{{{{\left| {{\mathbf{h}}_s^H{\mathbf{w}}} \right|}^2}}}{{\sigma _s^2}} \geqslant {\gamma _{th}} \hfill \\
  \quad\quad\quad\quad\;\;\;{\text{                  }}\left| {{{[{\mathbf{w}}]}_n}} \right| = \sqrt p ,n = 1,...,N \hfill \\ 
\end{gathered} .
\end{equation}
Similar to (\ref{eq10}), the original  problem in (\ref{eq28}) can be written as
\begin{equation}\label{eq29}
\begin{gathered}
  \mathop {\max }\limits_{\mathbf{w}} {\text{ }}\mathop {\min }\limits_{{{\mathbf{h}}_k} \in {\Psi _k}} {\text{  }}\frac{{1 + {{\mathbf{w}}^H}{{{\mathbf{\tilde h}}}_s}{\mathbf{\tilde h}}_s^H{\mathbf{w}}}}{{1 + \sum\limits_{k = 1}^K {\left( {{{\mathbf{w}}^H}{{{\mathbf{\tilde h}}}_k}{\mathbf{\tilde h}}_k^H{\mathbf{w}}} \right)} }} \hfill \\
  \;\;\;\;\;\;\;\;\;\;\;\;\;{\text{          s}}{\text{.t}}{\text{.    }}(11b),(11c) \hfill \\ 
\end{gathered} .
\end{equation}
Further, the optimization problem in (\ref{eq29}) is equivalent to the following alternative form
\begin{equation}\label{eq30}
\begin{gathered}
  \mathop {\min }\limits_{\mathbf{w}} {\text{   }}\frac{{1 + \sum\limits_{k = 1}^K {\left( {\mathop {\max }\limits_{{{\mathbf{h}}_k} \in {\Psi _k}} \left( {{{\mathbf{w}}^H}{{{\mathbf{\tilde h}}}_k}{\mathbf{\tilde h}}_k^H{\mathbf{w}}} \right)} \right)} }}{{1 + {{\mathbf{w}}^H}{{{\mathbf{\tilde h}}}_s}{\mathbf{\tilde h}}_s^H{\mathbf{w}}}} \hfill \\
  {\text{  s}}{\text{.t}}{\text{.   }}(11b),(11c) \hfill \\ 
\end{gathered} .
\end{equation}
With the help of  log-sum-exp function, we can obtain
\begin{equation}\label{eq31}
\begin{aligned}
\mathop {\max }\limits_{{{\mathbf{h}}_k} \in {\Psi _k}} {\text{ }}{{\mathbf{w}}^H}{{\mathbf{\tilde h}}_k}{\mathbf{\tilde h}}_k^H{\mathbf{w}} &= {\text{ }}\mathop {\max }\limits_{{{\mathbf{h}}_k} \in {\Lambda _k}} {\text{ }}{{\mathbf{w}}^H}{{\mathbf{\tilde h}}_k}{\mathbf{\tilde h}}_k^H{\mathbf{w}} \\
&\approx {\beta ^{ \!-\! 1}}\ln \!\!\left(\!\! {\sum\limits_{i = 0}^{{M_1} \!-\! 1} {\sum\limits_{j = 0}^{{M_2} \!-\! 1} {{e^{\beta \left( {{{\mathbf{w}}^H}{\mathbf{\tilde h}}_k^{\left( {i,j} \right)}{\mathbf{\tilde h}}{{_k^{\left( {i,j} \right)}}^H}{\mathbf{w}}} \right)}}} } } \!\!\right)\!\!
\end{aligned}.
\end{equation}
Therefore, the optimization  problem  (\ref{eq30}) can be converted to
\begin{equation}\label{eq32}
\begin{gathered}
  \mathop {\min }\limits_{\mathbf{w}} {\text{   }}\frac{{1 \!+\! \sum\limits_{k = 1}^K {\left( {{\beta ^{ \!-\! 1}}\ln \left( {\sum\limits_{i = 0}^{{M_1} \!-\! 1} {\sum\limits_{j = 0}^{{M_2} \!-\! 1} {{e^{\beta \left( {{{\mathbf{w}}^H}{\mathbf{\tilde h}}_k^{\left( {i,j} \right)}{\mathbf{\tilde h}}{{_k^{\left( {i,j} \right)}}^H}{\mathbf{w}}} \right)}}} } } \right)} \right)} }}{{1 + {{\mathbf{w}}^H}{{{\mathbf{\tilde h}}}_s}{\mathbf{\tilde h}}_s^H{\mathbf{w}}}} \hfill \\
  {\text{ s}}{\text{.t}}{\text{.   }}(11b),(11c) \hfill \\ 
\end{gathered} .
\end{equation}
Similar to the optimization problem in (\ref{eq18}), the objective function of this problem is non-convex and in the form of fractions, which can still be solved by the Dinkelbach method. The  corresponding sub-problem is given by
\begin{equation}\label{eq33}
\begin{gathered}
\mathop {\min }\limits_{\mathbf{w}} {\text{   }}1 + \sum\limits_{k = 1}^K {\left( {{\beta ^{ - 1}}\ln \left( {\sum\limits_{i = 0}^{{M_1} - 1} {\sum\limits_{j = 0}^{{M_2} - 1} {{e^{\beta \left( {{{\mathbf{w}}^H}{\mathbf{\tilde h}}_k^{\left( {i,j} \right)}{\mathbf{\tilde h}}{{_k^{\left( {i,j} \right)}}^H}{\mathbf{w}}} \right)}}} } } \right)} \right)}  \hfill \\
{\text{                    }} \quad\quad\quad\quad\quad\;\;\;\;\;\;\;- \eta \left( {1 + {{\mathbf{w}}^H}{{{\mathbf{\tilde h}}}_s}{\mathbf{\tilde h}}_s^H{\mathbf{w}}} \right) \hfill \\
\;\;{\text{s}}{\text{.t}}{\text{.   }}(11b),(11c) \hfill \\ 
\end{gathered}. 
\end{equation}
The optimization problem in (\ref{eq33}) has the same form as (\ref{eq19}), and thus can be solved by the non-convex ADMM algorithm framework, which is omitted here due to space limitation.
\begin{algorithm}
	\normalsize
	\caption{The proposed robust BF scheme.}
	\LinesNumbered 
	\KwIn{$\left\{ {{{\bf{h}}_s},{{\bf{h}}_k},p,{\gamma _{th}}} \right\}$}
	Initialize ${{\bf{w}}^{\!(\!0\!)\!}}$ as a feasible solution of    (\ref{eq18})  and  ${\eta ^{\!(\!0\!)\!}} \!\!=\!\! 0$\;
	Set the iteration times $t = 0$   and the calculation accuracy $\varepsilon $, $\delta $\;
	\Repeat{$\left| {{\eta ^{(t)}} - {\eta ^{(t - 1)}}} \right| \le \varepsilon$}{
		${\eta ^{(t + 1)}} \!\!=\!\! \frac{{{\beta ^{ - 1}}\ln \left( {\sum\limits_{k = 1}^K {\sum\limits_{i = 0}^{{M_1} - 1} {\sum\limits_{j = 0}^{{M_2} - 1} {{e^{\beta \left( {1 + {{\bf{w}}^{(t)}}^H{\bf{\tilde h}}_k^{\left( {i,j} \right)}{\bf{\tilde h}}{{_k^{\left( {i,j} \right)}}^H}{{\bf{w}}^{(t)}}} \right)}}} } } } \right)}}{{1 + {{\bf{w}}^{(t)}}^H{{{\bf{\tilde h}}}_s}{\bf{\tilde h}}_s^H{{\bf{w}}^{(t)}}}}$\;
		Initialize \!\!$\left\{ {{{{\bf{\tilde w}}}^{\!\!(0)\!\!}},{{\bf{x}}^{\!\!(0)\!\!}}} \right\}$\!\! as a feasible solution of  \!\! (\ref{eq20})\!\! \;
		Choose the Lagrangian multiplier ${{\bf{v}}^{(0)}}$, the penalty factor $\rho $ and set the iteration times $l = 0$\;
		\Repeat{$\left\| {{{{\bf{\tilde w}}}^{(l)}} - {{\bf{x}}^{(l)}}} \right\| \le \delta$}{
			Update   ${{\bf{x}}^{(l + 1)}}$ by solving sub-problem (\ref{eq22})\;
			Update ${{\bf{\tilde w}}^{(l + 1)}}$ by solving sub-problem (\ref{eq23})\;
			Update    ${{\bf{v}}^{(l + 1)}}$ by (\ref{eq24})\;
			Set $l = l + 1$\;
		}
		Update  ${{\bf{w}}^{(t + 1)}} = {{\bf{\tilde w}}^{(l)}}$\;
		Set $t = t + 1$\;
	}
	\KwOut{robust BF weight vector ${{\bf{w}}^{(t)}}$. }
\end{algorithm}
\section{Numerical Results}
In this section, we provide simulation results to verify the effectiveness and superiority of our proposed scheme. The MRT BF and the non-robust BF schemes are taken as benchmarks for the sake of  comparison. Unless otherwise indicated, we set the number of beams $N$=7, the number of Eves $K$=3, the QoS threshold of LU ${\gamma _{th}\!\!=\!\!5}$, the edge length of the uncertain area as 100 Km, and the approximation parameter $\beta {\text{ \!\!=\!\! 100}}$. The other system parameters are set by  referring to [\cite{references3}][\cite{references17}].

\vspace{-0.1cm}
\begin{figure*}
\centering
\begin{minipage}{0.32\textwidth}
\includegraphics[width=2.5in,height=2in]{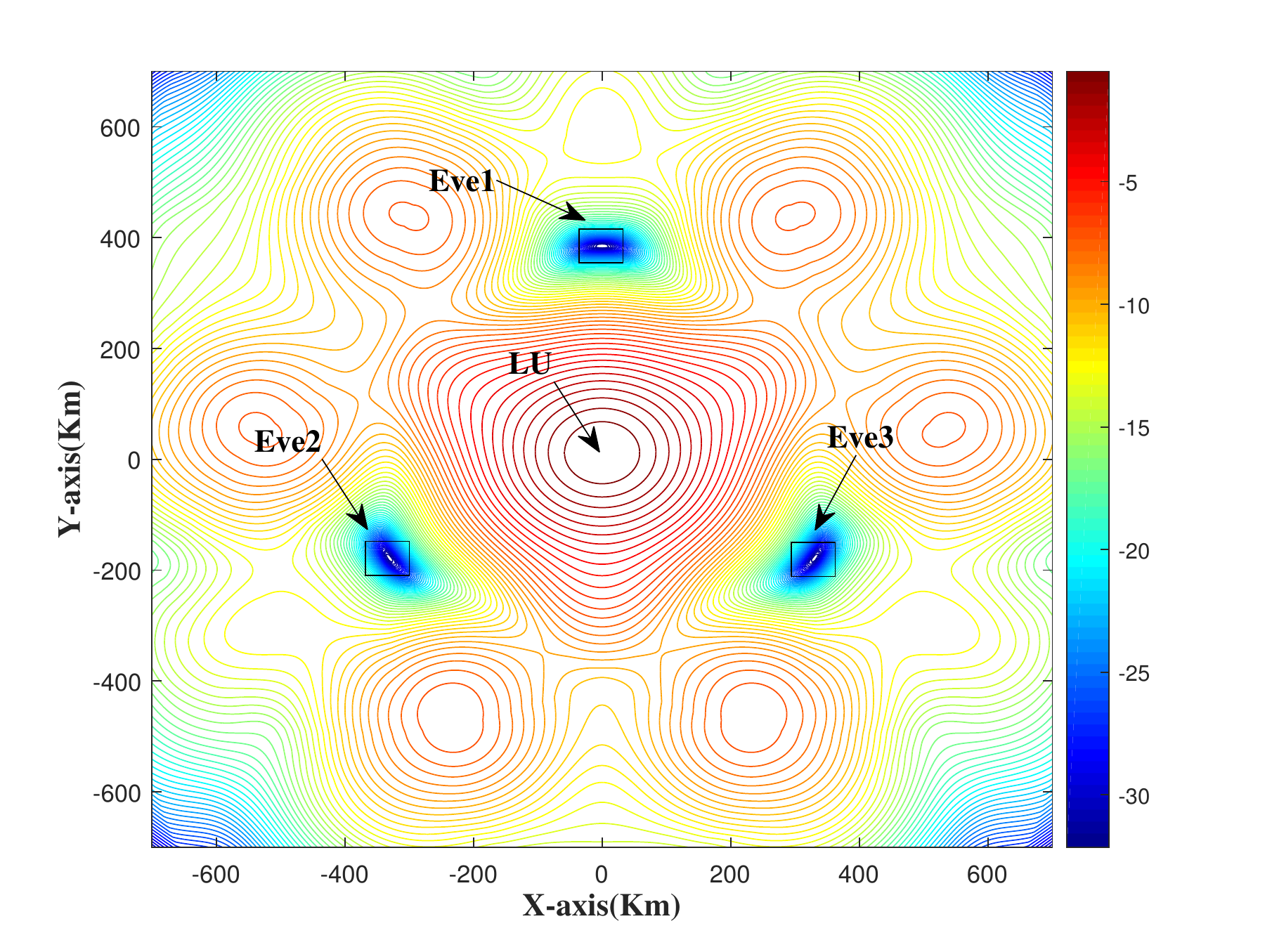}
\caption{Beampattern for  UE }
\label{fig:13averageLostTime}
\end{minipage}
\begin{minipage}{0.32\textwidth}
\includegraphics[width=2.3in,height=2in]{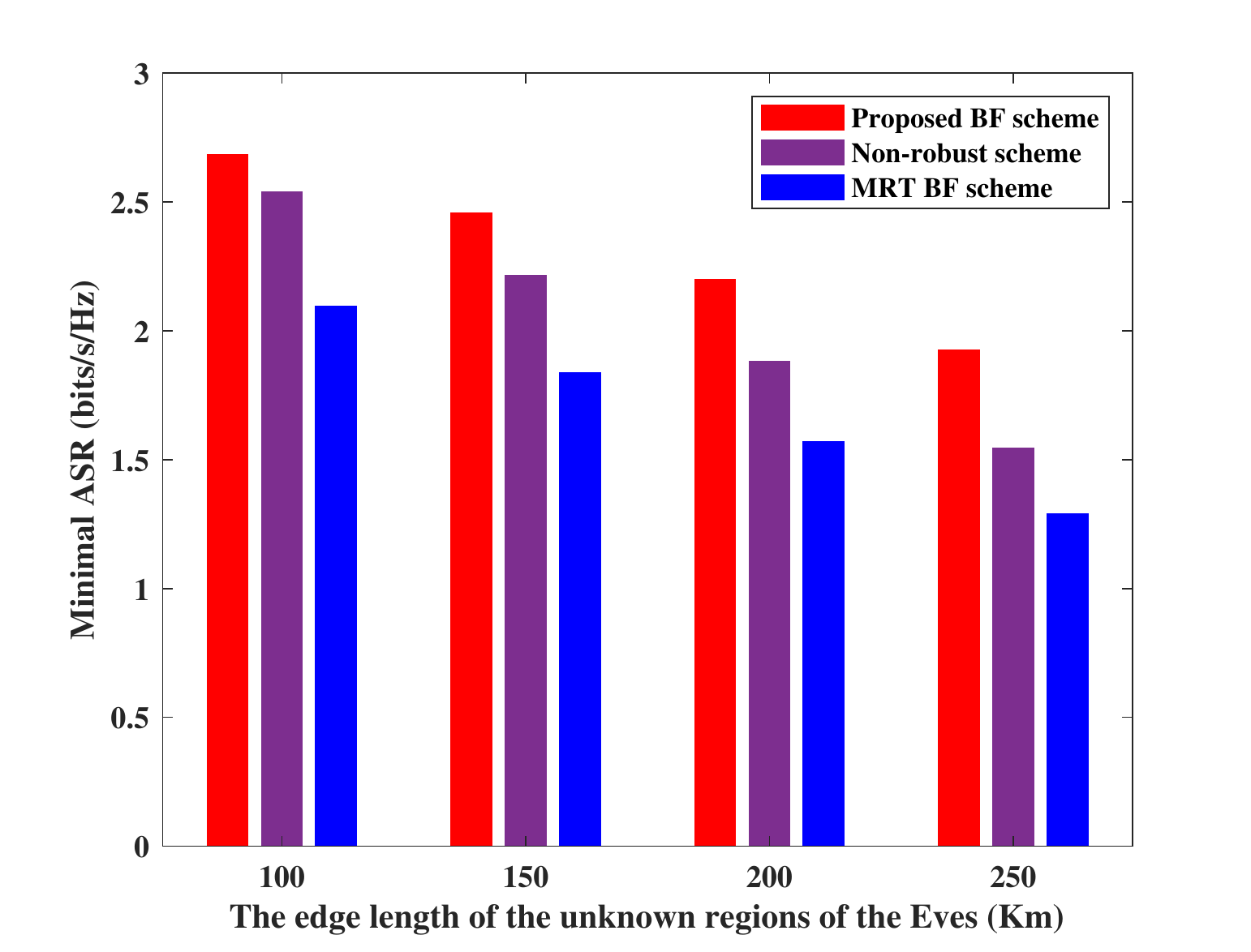}
\caption{Minimal ASR versus  the edge length of the unknown regions of the Eves}
\label{fig.3 Minimal ASR versus the edge length of the uncertain region}
\end{minipage}
\begin{minipage}{0.32\textwidth}
	\includegraphics[width=2.3in,height=2in]{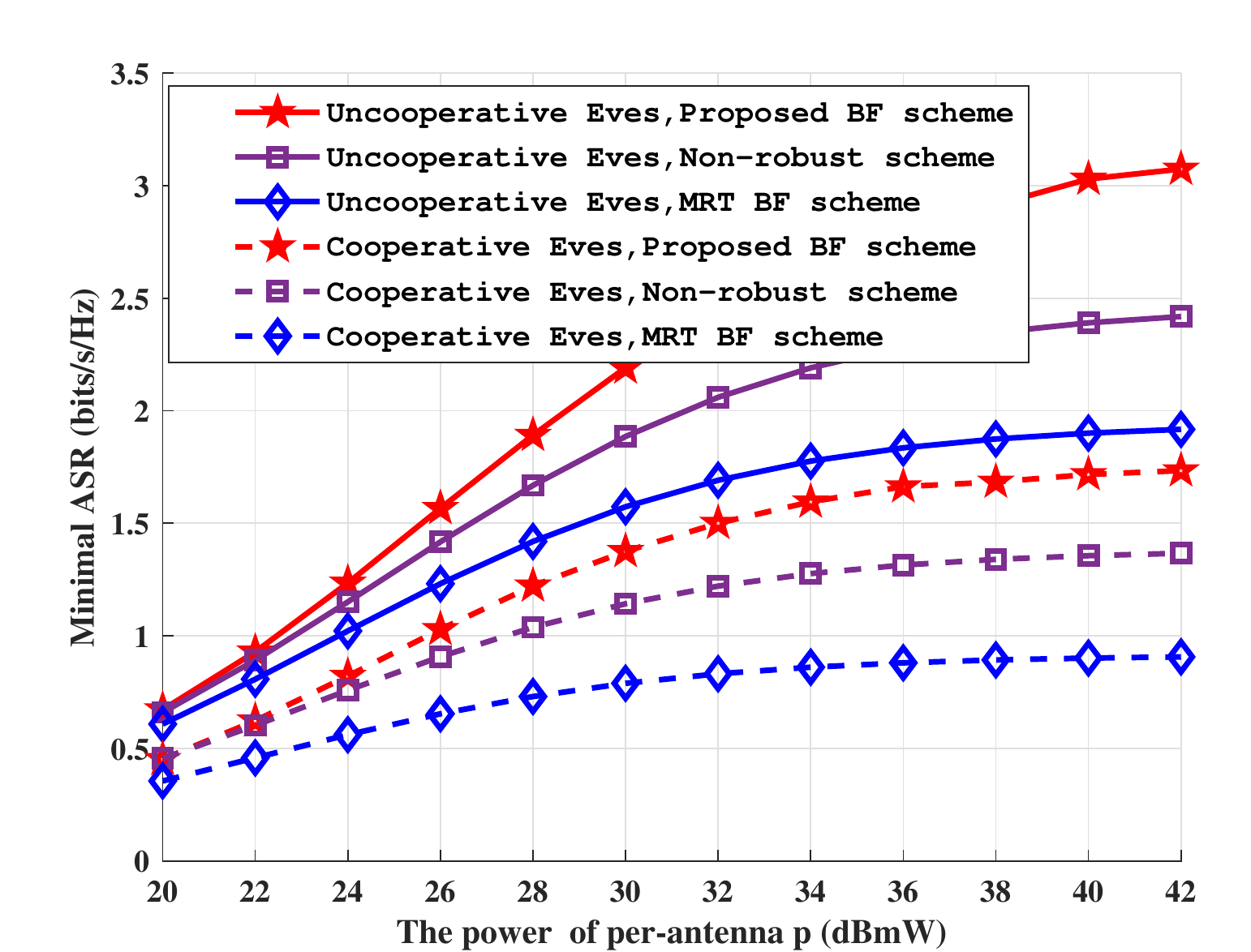}
	\caption{Minimal ASR versus the power  of per-antenna}
	\label{fig.3Minimal ASR versus the power of per-antenna.}
\end{minipage}
\end{figure*}
Fig. 1 presents the beampattern for the case of UE. It can be found that the maximal direction of beamforming points to LU, while three nulls are generated in the estimated region of Eves, respectively. It not only enhances the signal power of the LU but also inhibits the signal quality of the Eves.

The influence of the edge length of the unknown regions of the Eves on minimal ASR for the case of UE is depicted in Fig. 2. The per-antenna transmit power is assumed as 30 dBmW, it can be found that the minimum ASR of LU decreases as the edge length of the unknown regions of the Eves increases. Fig. 3 gives the minimal ASR versus the transmit power budget of the per-antenna for three BF schemes by assuming that the edge length of the unknown regions of the Eves is 200 Km. It can be observed that our proposed robust BF scheme is superior to the MRT BF scheme and the non-robust BF scheme in both cases of UE and CE. The reason is that the MRT scheme only focuses on the maximum SNR of the LU, without considering the ASR, and the non-robust BF scheme does not take the  unknown regions of the Eves into consideration. 
\begin{table}[h]
	\small
	\caption{Comparison of the CPU running times (seconds)}\label{Table1parameters}
	\centering
	\begin{tabular}{|c|c|c|}% {ccc} 表示各列元素对齐方式，left-l,right-r,center-c
		\hline
		Number of Eves & SDR CPU Time & Proposed CPU Time\\% \hline 在此行下面画一横线 % \\ 表示重新开始一行 % & 表示列的分隔线
		\hline
		1  & 1.012 & 0.0147 \\
		\hline
		2  & 1.406 & 0.0445 \\
		\hline
		3  & 1.911 & 0.0759 \\
		\hline
		4  & 2.517 & 0.1733 \\
		\hline
	\end{tabular}
\end{table}

To confirm the computational advantage of the proposed algorithm, the SDR-based method is adopted for comparison. The CPU running times of the two approaches, with different numbers of Eves, are listed in Table I. It can be observed that the proposed algorithm runs much faster than the SDR method.
\vspace{-0.2cm}
\section{Conclusion}
\vspace{-0.1cm}
This paper has investigated the PLS in the downlink of a multibeam satellite system. Specifically, we first formulated a optimization problem to maximize the worst-case ASR of the LU. Since the original problem is nonconvex and intractable, we adopted log-sum-exp function to approximate the objective function, and then exploited ADMM combined with Dinkelbach method to obtain the BF weight vector. Finally, simulation results have confirmed the effectiveness and superiority of the proposed robust BF scheme. In our future works, we will consider other effects, such as hardware impairments in BF design to enhance security in multibeam satellite systems.

\ifCLASSOPTIONcaptionsoff
  \newpage
\fi

%\scriptsize
\footnotesize
%\small
\bibliographystyle{IEEEtran} %(plain可以替换为你所投杂志的BTS文件名，其控制引用参考文献格式）
\bibliography{IEEEabrv,IEEEexample} % mybib是Endnote导出的文件名，即你的文献库bib格式

\end{document}